\documentclass[11pt]{article}

\usepackage{amsfonts,amsmath,amssymb,mathrsfs}
\usepackage{hyperref}
\usepackage{color}
\usepackage{graphicx}  
\usepackage{dcolumn}   
\usepackage{bm}        
\usepackage[english]{babel}
\def\a{\alpha}

\def\r{\rho}
\def\s{\sigma}
\def\t{\tau}
\def\m{\mu}
\def\n{\nu}
\def\k{\kappa}
\def\th{\theta}
\def\g{\gamma}\def\G{\Gamma}
\def\L{\Lambda}\def\l{V}
\def\D{\Delta}
\def\la{\langle}
\def\ra{\rangle}
\def\o{\omega}\def\O{\Omega}
\def\d{\delta}
\def\p{\partial}

\def\oxthree{{\cal O}(x^3) }

\def\half{\textstyle{\frac{1}{2}}}

\def\bdoc{\begin{document}}
\def\edoc{\end{document}}
\def\bea{\begin{equation}}
\def\eea{\end{equation}}

\def\beq{\begin{eqnarray}}
\def\eeq{\end{eqnarray}}
\def\ben{\begin{enumerate}}
\def\een{\end{enumerate}}
\def\la{\langle}
\def\ra{\rangle}
\def\a{\alpha}
\def\g{\gamma}\def\G{\Gamma}
\def\d{\delta}\def\D{\Delta}
\def\e{\epsilon}
\def\z{\zeta}

\def\th{\theta}
\def\k{\kappa}
\def\l{\lambda}
\def\m{\mu}
\def\n{\nu}
\def\o{\omega}
\def\p{\pi}
\def\r{\rho}
\def\s{\sigma}
\def\t{\tau}
\def\L{{\cal L}}
\def\S{\Sigma }
\def\gsim{\; \raisebox{-.8ex}{$\stackrel{\textstyle >}{\sim}$}\;}
\def\lsim{\; \raisebox{-.8ex}{$\stackrel{\textstyle <}{\sim}$}\;}
\def\gtrsim{\gsim}
\def\lessim{\lsim}
\def\loc{{\rm local}}
\def\vm{v_{\rm max}}
\def\bh{\bar{h}}
\def\del{\partial}
\def\nab{\nabla}
\def\half{{\textstyle{\frac{1}{2}}}}
\def\fourth{{\textstyle{\frac{1}{4}}}}

\def\bD{{\bf D}}
\def\bE{{\bf E}}
\def\bF{{\bf F}}
\def\bB{{\bf B}}
\def\bP{{\bf P}}
\def\bV{{\bf v}}
\def\bv{{\bf v}}
\def\bx{{\bf x}}
\def\by{{\bf y}}
\def\bz{{\bf z}}
\def\ba{{\bf a}}
\def\bd{{\bf d}}
\def\bs{{\bf s}}
\def\bn{{\bf n}}
\def\bp{{\bf p}}

\def\O{\Omega}

\def\br{{\bf r}}
\def\bnab{{\bf \nab}}

\def\tE{\tilde{E}}
\def\tL{\tilde{L}}
\def\Horava{Ho\v{r}ava }

\def\oxtwo{\mathscr{O}\left(x^2\right)}
\def\oxthree{\mathscr{O}\left(x^3\right)}
\def\oxfour{\mathscr{O}\left(x^4\right)}
\def\oxfive{\mathscr{O}\left(x^5\right)}

\def\ph{\phantom}

\def\LL{Lanczos-Lovelock}

\begin{document}

\title{Indistinguishability of thermal and quantum fluctuations}
\author{Sanved Kolekar\footnote{sanved@iucaa.ernet.in}, T. Padmanabhan\footnote{paddy@iucaa.ernet.in}\\
IUCAA, Pune University Campus, Ganeshkhind,\\
Pune 411007, India.}
\date{\today}
\maketitle
\begin{abstract}
The existence of Davies-Unruh temperature in a uniformly accelerated frame shows that quantum fluctuations of the inertial \textit{vacuum} state appears as thermal fluctuations in the accelerated frame. Hence thermodynamic experiments cannot distinguish between phenomena occurring in a thermal bath of temperature $T$ in the inertial frame from those in a frame accelerating through inertial vacuum  with the acceleration $a=2\pi T$.  We show  that this indisguishability between quantum fluctuations and thermal fluctuations goes far beyond the  fluctuations in the vacuum state. We show by an exact calculation, that the reduced density matrix for a uniformly accelerated observer when the quantum field is in a \textit{thermal} state of temperature $ T^\prime$, is symmetric between acceleration temperature $T = a/(2\pi)$ and the thermal bath temperature $T^\prime$. Thus thermal phenomena cannot distinguish whether (i) one is accelerating with $a = 2\pi T$ through a bath of temperature $T^\prime$ or (ii) accelerating with $a=2\pi T^\prime$ through a bath of temperature $T$. This shows that thermal and quantum fluctuations in an accelerated frame affect the observer in a symmetric manner. The implications are discussed.
\end{abstract}
\maketitle

It is well known that the inertial vacuum state of a quantum field appears to an accelerated observer as  \cite{Davies, Unruh} a thermal state with temperature  $T$ related to the magnitude of its acceleration $a$, as $T =a/2\pi$. We use units with $c=k_B = \hslash = 1$. More precisely, the reduced density matrix,
obtained by tracing over degrees of freedom hidden behind the Rindler horizon, when the field is in the inertial vacuum state, has the form of a thermal density matrix with temperature $T= a/2\pi$. Hence the appearance of vacuum fluctuations to a  uniformly accelerated observer is indistinguishable  from the thermal fluctuations seen by a stationary observer in a \textit{real} thermal bath. This point of view is also supported by the fact that  a two level quantum system undergoing uniform acceleration and coupled to the quantum field in its Minkowski vacuum state, will reach a level population \cite{unruhdet} that is identical to that of a detector immersed in a thermal bath after it has reached equilibrium.

Such results regarding the indistinguishability of thermal and vacuum fluctuations are based on studying  the inertial\textit{ vacuum }state of the quantum field. In other words, the fluctuations present in the inertial frame are purely those of the inertial vacuum which, from the perspective of an accelerated observer, appears as thermal fluctuations. However, it is not clear what is the relationship between the quantum fluctuations and the thermal fluctuations for the accelerated observer when the latter are \textit{also} present in the inertial frame. To emphasize this point, let  us consider a real thermal bath at temperature $T^\prime$ in the inertial vacuum which will introduce genuine thermal fluctuations in the inertial frame. Consider now an observer moving through this thermal bath with an acceleration $a$ corresponding to a Davies-Unruh temperature  $T=a/(2\pi)$. The thermal phenomena seen by such an observer can be described by a suitable theoretical construct, say, a density matrix $\rho (T, T^\prime)$ which could depend on both $T$ and $ T^\prime$ (and on possibly other variables which are irrelevant to us). We use the convention that, in the density matrix $\rho(T_1,T_2)$ the first argument denotes the acceleration temperature and the second one denotes the temperature of a thermal bath as measured in the inertial frame. The Davies-Unruh effect tantamounts to the statement that 
\begin{equation}
 \rho (0, T) = \rho (T,0)
\end{equation} 
This prompts us to ask the question: \textit{Does the indistinguishability extend to situations when both thermal and quantum fluctuations are simultaneously present?} That is, can we prove that 
\begin{equation}
 \rho (T, T^\prime) = \rho ( T^\prime, T)
 \label{equi}
\end{equation} 
when both $T$ and $T^\prime$ are non-zero?
Such an equality would imply that an observer accelerating through a thermal bath of temperature $ T^\prime$ with an acceleration corresponding to temperature $T$ will experience the same thermal phenomena as
an observer accelerating through a thermal bath of temperature $ T$ with an acceleration corresponding to temperature $ T^\prime$. 

The key result of this paper is the rigorous proof of the above result in Eq.~(\ref{equi}) for a thermal bath given in Eq.~(\ref{thermalbath}) thereby escalating the equivalence between quantum and thermal phenomena to a new level. We do this  by computing 
the reduced density matrix for a uniformly accelerated observer with acceleration $a$ when the quantum field is  in a thermal bath with temperature $(1/ \beta^\prime)$.  The thermal bath considered consists of the Unruh particles (the reason for such a choice over a thermal bath of Minkowski particles is explained
below). We calculate the resultant density matrix in the usual way of tracing over the degrees of freedom hidden behind the Rindler horizon. We obtain an exact expression without any approximation which by itself is an interesting result and first of its kind for a thermal bath.

We note that there have been previous works (see for example ref. \cite{accthermalbath, accthermalbath2}) which calculated the \textit{spectrum} of  particles detected by uniformly accelerated observers in an inertial thermal bath by computing  the Bogoliubov co-efficients  or the excitation rate of the Unruh Dewitt detector. However, these methods only tell us the number density of Rindler particles detected whereas to make a definite statement about the indisguishability of quantum and thermal fluctuations, one must have a complete knowledge of the system which is precisely provided by the reduced density matrix. Furthermore, the calculation using detectors coupled to the thermal bath involves a prescription of averaging over different excitation rates with appropriate thermal weightages and is not a derivation from first principles. The density matrix formalism  works without any approximations and the result we  obtain is exact and  from first principles. In the context of relativistic quantum
entanglement a similar computation was done, although with completely
different purposes, using the covariant matrix formalism to understand the
effect of the initial temperature of two modes on the final entanglement after
the modes are two-mode-squeezed \cite{brus}.

We will begin by  summarizing the procedure involved  in calculating the reduced density matrix. We consider a free scalar field described by  the Klein Gordon field equation. Let the \textit{state} of the scalar field with respect to the inertial observer be described by a density matrix $\rho_{M} $ which - in general - can represent a mixed state or a thermal bath. 
One can express  $\rho_M $ in terms of any basis eigenstates which are a complete set of basis vectors. The eigenstates of the inertial Hamiltonian and the eigenstates corresponding to the left and right Rindler Hamiltonian together, are such complete set of basis vectors. Let us denote by $|_L n \rangle$ and $|_R n \rangle$  the left and right Rindler Hamiltonian eigenstates respectively and $(b^\dagger_{(L)k},b_{(L)k})$ and $(b^\dagger_{(R)k},b_{(R)k})$  be the creation and annihilation operators corresponding to the left and right Rindler wedges respectively. An observer confined to the right Rindler wedge $X>|T|$ will have all her physical observables made purely out of $(b^\dagger_{(R)k},b_{(R)k})$ and they will be independent of $(b^\dagger_{(L)k},b_{(L)k})$. Let
${\cal O}(b^\dagger_{(R)k},b_{(R)k})$ be any such observable. Then expectation value of ${\cal O}$ for a given density matrix $\rho_M$ is
\begin{eqnarray}
\langle {\cal O} \rangle &=& Tr(\rho_M {\cal O}) \nonumber \\
&=& \sum_{p,q}\langle p|_R \otimes \langle q|_L \; (\rho_M {\cal O})\; |_L q\rangle  \otimes |_R p\rangle \nonumber \\
&=& \sum_{p}\langle p|_R \; (\rho_r {\cal O}) \; |_R p\rangle
\label{expectation}
\end{eqnarray}
where in the last step we have used the fact that ${\cal O}$ is independent of $(b^\dagger_{(L)k},b_{(L)k})$ and defined the reduced density matrix $\rho_r$ as 
 \begin{eqnarray}
\rho_r &=& \sum_{q}\langle q|_L \; \rho_M \; |_L q\rangle
\end{eqnarray}
It is evident from Eq.[\ref{expectation}] that the reduced density matrix $\rho_r$ is the one relevant for our  purpose.

It is known that when $\rho_M = | 0_M\rangle \langle 0_M|$ (where  $|0_M\rangle$ is the Minkowski vacuum state) the reduced density matrix $\rho_r$ corresponding to the right (or left) Rindler observer is a thermal density matrix with temperature $\beta^{-1} = a/2\pi$. It can be expressed in the following form
\begin{eqnarray}
\rho_{unruh} &=& \prod_k C_k^2 \sum_n e^{-n \beta \omega_k} |_R n_k\rangle \langle n_k|_R \nonumber \\
&=& \prod_k C_k^2 \sum_n e^{-n  \beta \omega_k} \frac{(b^\dagger_{(R)k})^n}{n!} | 0_R\rangle \langle 0_R| (b_{(R)k})^n
\label{Unruhbath}
\end{eqnarray}
corresponding to the standard Davies-Unruh effect. The expectation value of the number operator $(b^\dagger_{(R)k}b_{(R)k})$ for the above thermal density matrix leads to the Planckian distribution
\begin{eqnarray}
\langle {\cal N} \rangle &=& Tr[\rho_{unruh}b^\dagger_{(R)k}b_{(R)k}] \nonumber \\
&=& \frac{1}{\left( e^{\beta \omega_k} -1 \right)}
\end{eqnarray}
 
In our case, we want to obtain the form of the reduced density matrix $\rho_r$ when the scalar field is described by an  thermal density matrix $\rho_{th} = \exp(-{ \beta^\prime {\cal H}})$ with temperature given by $(1/ \beta^\prime)$.
Ideally, one would like to start with $\rho_{th} = \exp(-{ \beta^\prime {\cal H}})$ with the Hamiltonian being the sum of  ${\cal H}_k = \omega_k a^\dagger_{k}a_{k}$ where $a^\dagger_{k}$ and $a_{k}$ are the creation and annihilation operators respectively defined with respect to the inertial plane wave modes. However, as is well known,   $a_{k} f\left(b_{k_1},b_{k_2},b_{k_3},...,b^\dagger_{k_1},b^\dagger_{k_2},...\right)$ is a function of the Rindler creation and annihilation operators of all frequencies, and hence, an inertial plane wave mode of frequency $\omega_k$ gets mixed with Rindler plane wave modes of all frequencies. So 
calculating the reduced density matrix in an exact closed form is algebraically untractable.

We shall avoid this problem of mode mixing by exploiting  a trick originally devised by Unruh. This involves working with the  so called Unruh modes $U^{(1)}_k, U^{(2)}_k$ \cite{unruhdet} which are linear combination of the Minkowski plane wave modes such that negative and positive frequencies do not mix. Due to this, the vacuum state corresponding to the plane wave mode solutions and the Unruh modes remains the same. Further, the Unruh mode of frequency $k$ maps to Rindler modes only with frequencies $k$ and $-k$, which is the key aspect required to avoid the mode mixing problem. This significantly simplifies the calculation and, as we shall demonstrate below, allows us to obtain an exact closed expression for $\rho_r$.  (We also point out later that working with Unruh particles rather than with Minkowski particles is advantageous in dealing with non-stationarity issues.)

There are two sets of creation and annihilation operators $d^\dagger_{1k}, d_{1k}$ and $d^\dagger_{2k}, d_{2k}$ corresponding to the two Unruh modes $(U^{(1)}_k, U^{\dagger(1)}_k)$ and $(U^{(2)}_k, U^{\dagger(2)}_k)$ respectively. We begin by defining $ \rho_{th} = e^{- \beta^\prime {\cal H}}$ where ${\cal H} = \sum_k \omega_k d^\dagger_{1k}d_{1k} $ to be a thermal density matrix describing a thermal bath of Unruh particles. One should note that this Hamiltonian ${\cal H} = \sum_k \omega_k d^\dagger_{1k}d_{1k} $ is not the original
Hamiltonian for the scalar field defined in terms of the usual Minkowski
creation and annihilation operators and hence lacks a natural interpretation
from an inertial observer’s perspective as is well known in the literature.
Nonetheless, for our present purpose, we only require the thermal
state to be populated by particles through excitations over the standard
Minkowski/inertial vacuum. We consider a thermal bath of Unruh particles
created by $d_{1k}^\dagger$ acting on the Minkowski vacuum with the interpretation of $1/\beta^{\prime}$ as the temperature of the bath. These Unruh particles are then, of
course, different from the standard Minkowski particles but both belong to
excited states of the same vacuum. Using the basis representation of the eigenstates of the Hamiltonian operator ${\cal H}_{inertial} = \sum_k \omega_k a^\dagger_{k}a_{k} $, we define $\rho_{th}$ to be  
\begin{eqnarray}
\rho_{th} = \prod_k C_k^2 \sum_m e^{-m  \beta^\prime \omega_k} \frac{(d^\dagger_{1k})^m}{\sqrt{m!}} | 0_M\rangle \langle 0_M| \frac{(d_{1k})^m}{\sqrt{m!}}
\label{thermalbath}
\end{eqnarray}
where $C_k^2$ is the normalization constant. 
A  comparison of the form of the density matrix in Eq.[\ref{thermalbath}] with that of Eq.[\ref{Unruhbath}], shows that $ \rho_{th} $  is  a thermal density matrix with ${\cal H}$ playing the role the \textit{effective} Hamiltonian for the Unruh particles corresponding to the operators $d^\dagger_{1k}, d_{1k}$. 
 We could have chosen to work with a thermal bath of $d^\dagger_{2k}, d_{2k}$ particles 
 instead of a thermal bath of $d^\dagger_{1k}, d_{1k}$ particles  without any loss of generality. However, one must then take care to 
 trace over appropriate set of mode, left or right, in the two contexts. This is because, by construction, the two set of Unruh modes are asymmetric in terms of the left and right Rindler modes. For, example  $d^\dagger_{1k}$ is a function of only the left Rindler creation and the right Rindler annihilation operators and hence, for a thermal bath of $d^\dagger_{1k}, d_{1k}$ particles, one must trace over the right modes.  Similarly, for the thermal bath of  $d^\dagger_{2k}, d_{2k}$ particles, one must trace over the left Rindler modes. 
 Further, it is easy to verify that  $\left[H_{d_1}, \rho_{th} \right] = 0 = \left[H_{b_L}, \rho_{th} \right]$ where $H_{b_L}$ and $H_{d_1}$ are the left Rindler Hamiltonian and Hamiltonian corresponding to the Unruh mode $(U^{(1)}_k)$ respectively. The vanishing commutators show that $\rho_{th}$ is stationary with respect to both $H_{b_L}$ and $H_{d_1}$ and hence the comparison of $T$ and $T'$ can be performed at any instant of time. Thus, the choice of the thermal bath in terms of the Unruh particles rather than the Minkowski particles avoids the well known non-stationarity issues which arise in the latter case \cite{accthermalbath2}.

We can now proceed to determine the corresponding reduced density matrix $\rho_r$. Our objective will be to take the trace of $\rho_{th}$ with respect to the right Rindler states.
The calculation is straightforward but lengthy and the key steps are indicated in Appendix A.
The final result can be expressed as:
\begin{eqnarray}
\rho_{r} &=& \prod_k C_k^2 \sum_{p} \left[1+e^{- ( \beta^\prime -\beta) \omega_k}  (1 - e^{- \beta \omega_k})\right]^{p} \nonumber \\
&& \times  e^{ -p\beta \omega_k} \; |_L p\rangle \langle p |_L \nonumber \\
\label{leftreducedM}
\end{eqnarray}
which is obtained  without any approximation and is exact. The normalization constant can be found by imposing $Tr(\rho_r) =1$ leading to  $C_k^2 = (1 - e^{- \beta \omega_k})(1 - e^{- \beta^\prime \omega_k})$. The expectation value of the number operator is
\begin{eqnarray}
\langle {\cal N} \rangle &=& Tr[\rho_{r}b^\dagger_{(R)k}b_{(R)k}] \nonumber \\
&=& \frac{1}{\left( e^{\beta \omega_k} -1 \right)} +\frac{1}{\left( e^{ \beta^\prime \omega_k} -1 \right)} + \frac{1}{\left( e^{ \beta^\prime \omega_k} -1 \right)\left( e^{\beta \omega_k} -1 \right)} \nonumber \\
\label{leftexpectation}
\end{eqnarray}
We note that an expression similar to Eq.[\ref{leftexpectation}] was obtained in \cite{brus}, but in
a completely different context using the covariant matrix formalism. However,
\cite{brus} neither mentions anything about the symmetry in temperatures nor
does it attempt to discuss the significance of it, most probably because the
authors have missed it due to their focus on completely different aspects.
(We thank an anonymous referee who brought this work to our attention.)
Further, it should also be mentioned that authors of reference \cite{brus} work with
two different initial thermal baths. We have considered the initial thermal
bath comprising of one (any one) of the two types of Unruh particles as
mentioned above whereas \cite{brus} considers both type of Unruh particles in thermal
equilibrium at same temperature. Due to this reason, the Eq.(15) in \cite{brus}
is similar to the above Eq.[\ref{leftexpectation}], but not quite the same.

We now identify $(Z_k^\beta)^{-1} = (1 - e^{- \beta \omega_k})$ to be the partition function of a thermal bath at temperature $\beta^{-1}$ corresponding to some $\rho_{thermal} = e^{-\beta {\cal H_k}}/ Z_k$ and similarly $(Z_k^{ \beta^\prime})^{-1} = (1 - e^{- \beta^\prime \omega_k})$ to be the partition functions of a thermal bath at temperature $(1/\beta^\prime)$. The form of $\rho_r$ in Eq.[\ref{leftreducedM}] in terms of these variables can then be simply written as  
\begin{eqnarray}
\rho_{r} &=& \prod_k (Z_k^\beta Z_k^{ \beta^\prime})^{-1} \sum_{p} \left[1 - (Z_k^\beta Z_k^{ \beta^\prime})^{-1} \right]^{p} \; |_L p\rangle \langle p |_L \nonumber \\
&=& \prod_k ({\bar Z_k(\beta, \beta^\prime)})^{-1} \sum_{p} \left[1 - ({\bar Z_k(\beta, \beta^\prime)})^{-1} \right]^{p} \; |_L p\rangle \langle p |_L \nonumber \\
\label{leftreducedM2}
\end{eqnarray}
where $ {\bar Z_k(\beta, \beta^\prime)} = Z_k^\beta Z_k^{ \beta^\prime}$ is the effective partition function of the reduced density matrix we are interested in.

\textit{We thus have a remarkable result: The effective partition function $ {\bar Z_k(\beta,  \beta^\prime)}$ is just the  product of the two thermal partition functions with temperatures $(1/\beta^\prime)$ and $ (1/\beta)$!} One can check that in the limit $ (\beta^\prime)^{-1} \rightarrow 0$ when there is no thermal bath, we have $Z_k^{ \beta^\prime} \rightarrow 1$ and hence we get back the Unruh effect, $\rho_r \rightarrow \rho_{unruh}$. Similarly, in the limit when the acceleration vanishes,  $ \beta^{-1} \rightarrow 0$, we have $Z_k^{ \beta} \rightarrow 1$ and hence we get initial thermal bath, $\rho_r \rightarrow \rho_{thermal}$.

There are several curious features about the result obtained  which are worth mentioning. To begin with, there have been suggestions in the literature \cite{Smolin} that quantum fluctuations and all statistical fluctuations, including thermal ones, are essentially identical, in the sense that both can be dissipative in nature, except in a globally flat spacetime, in which, the quantum fluctuations are conservative for  the inertial trajectories. Then, assuming a set of principles, Smolin  proposed that in all the other cases, (i.e., except for motion is along inertial trajectories in globally flat spacetimes) the quantum fluctuations should be indistinguishable from the statistical fluctuations. Our result resonates well with an equivalence of this kind though ours is  a specific and precise statement of the indistinguishability between thermal and vacuum fluctuations. Of course, if one performs, say, mechanical experiments which involve the measurement of the acceleration of the observer, then one can deduce the acceleration temperature $\beta^{-1}$ indirectly by making use of Unruh's result, $T = a/2\pi$. However, in the present context, the indistinguishability being discussed is purely within the thermodynamic domain of experiments in which the expectation value of any physical observable of the accelerated observer is calculated from $\rho_r$ according to Eq.[\ref{expectation}]. It may further be noted that some of the previous works \cite{Steeg} have claimed
to be able to distinguish between thermal and quantum fluctuations using
two detectors by considering the entanglement between them. However, a
careful reading of these works shows that the claim about two detectors
being able to distinguish (whereas a single detector cannot) between the
two kinds of fluctuations actually refers to the two kinds of fluctuations for
quantum systems in two different spacetimes. For e.g in \cite{Steeg}, the entanglement
between two detectors moving in a thermal bath in (i) Minkowski and
(ii) De-sitter are compared. Whereas, in the present paper, the two kinds of
fluctuations we have discussed are in terms of two temperature parameter $T$ and $T^\prime$ for the same quantum system in a given spacetime vis-a -vis, the
Minkowski spacetime. Hence these results cannot be compared directly with
our results.

Second,  while the result demonstrates the validity of Eq.~(\ref{equi}), it does not lead to any simple rule for combining the two temperatures. There is no simple way of defining an effective temperature for the resulting system. This is clear from the fact that $\rho_r$ is not thermal for the special case of $\beta = \beta^\prime$. One way to understand this non-thermality is in terms of the detailed
balance equation consisting of the induced probability to absorb or
emit particles which demonstrates thermality for an uniformly accelerated
detector moving in inertial vacuum fluctuations (see for example \cite{agullo}). In
the present case, when thermal fluctuations are additionally present, our result suggests that the delicate balance of the stimulated and absorption
probability rates gets modified thereby breaking the thermal equilibrium.
Such kind of emission processes are also evident when one inspects the expectation
value of the Rindler number operator as given in Eq.[\ref{leftexpectation}]. It has
the form $\langle {\cal N} \rangle = n_T + n_{T^\prime} + (n_{T})(n_{T^\prime})$ where $n_T$ and $n_{T^\prime}$ each represent
a Planckian distribution of particles at respective temperatures, $T$ and $T^\prime$. One can interpret the resultant expectation value as a stimulated emission
process wherein the presence of thermal bath $n_{T^\prime}$ stimulates an additional
emission of $n_T$ particles or vice-versa due to the symmetry in $T$ and $T^\prime$.

However, the symmetry in Eq.~(\ref{equi}) is  strengthened, in view of our result, to a stronger statement, viz., that the partition function $Z(T, T^\prime)$ is actually a \textit{product} function of the form $Z(T,  T^\prime) =f(T) f( T^\prime)$. This is in contrast to simpler forms of obtaining the symmetry in Eq.~(\ref{equi}) like $\rho(T,  T^\prime)$ just being a function of, say, $(T+ T^\prime)$ or $(T^2 + T^{\prime2})$ etc. The latter situation in fact arises in the context of a uniformly accelerated observer in the de Sitter spacetime \cite{deser} in which case, it is known that the effective temperature is the Pythagorean sum of the acceleration temperature  and de Sitter temperature. It would be interesting to study uniformly accelerated observers in other spacetimes with horizons to explore the connection with our result.

Finally, given the product structure of the full partition function $Z(\beta_1,\beta_2) = Z(\beta_1)Z(\beta_2)$, it is clear that $\ln Z(\beta_1,\beta_2)$ is additive. This, however, does \textit{not} imply simple additivity of thermodynamic variables because those require additivity of free energy $F = -\beta^{-1}\ln Z$ which cannot be defined without a natural notion of $\beta$ for the full system. The best we can do is to introduce a relation $\beta F = \beta_1 F_1 + \beta_2 F_2$ which is insufficient to define $\beta$ and $F$ individually. Many of these aspects are worth further investigation.

One must note that the symmetry of temperatures in the stationary
reduced density matrix of Eq.~(\ref{leftreducedM2}) is valid for a thermal bath of Unruh particles
given by Eq.~(\ref{thermalbath}) and may not hold in the natural case of a thermal
bath consisting of Minkowski particles due to the non-stationarity arguments
mentioned earlier. In the latter case, one would instead expect a time
dependent reduced density matrix and the relationship between $T$ and $T^\prime$ would be worth investigating

\textit{Acknowledgements:} We thank Jorma Louko for useful discussions and comments. TP's research is partially supported by J.C.Bose Research grant.

\section*{Appendix A}

We first express the Unruh operators in terms of the left and right Rindler operators as 
\begin{eqnarray}
d^\dagger_{1k} = \frac{\left[ b^\dagger_{(L)k} - {\bar q} \ b_{(R)(-k)} \right]}{{\bar p}\left(1 - {\bar q} \right)}
\label{d1intermsofb}
\end{eqnarray}
where ${\bar q} = \exp{(-\beta \omega_k/2)}$ and ${\bar p}^2 = \exp(\beta \omega_k/2)/[2\sinh{(\beta \omega_k/2)}]$. 
Further we use the fact that the Minkowski vacuum state can be expanded in terms of the left and right Rindler basis states as
\begin{eqnarray}
| 0_M\rangle = \prod_k A_k^2 \; \sum_n e^{-\frac{n}{2} \beta \omega_k} \;\; |_L n\rangle  \otimes |_R n\rangle
\label{minvacuum}
\end{eqnarray}
Using Eq.[\ref{minvacuum}], we can express the thermal density matrix in Eq.[\ref{thermalbath}] in terms of the left and right Rindler basis states as
\begin{eqnarray}
\rho_{th} & = & \prod_k C_k^2 A_k^2 \sum_{m,p,q} e^{-m \beta^\prime \omega_k}  e^{-\frac{(p+q}{2} \beta \omega_k} \nonumber \\
& &\times \frac{(d^\dagger_{1k})^m}{m!} |_L p\rangle  \otimes |_R p\rangle \langle q|_R \otimes \langle q|_L (d_{1k})^m
\label{thermalbath2}
\end{eqnarray}
Next we  binomially expand the right hand side of Eq.[\ref{d1intermsofb}] to get
\begin{eqnarray}
\left( d^\dagger_{1k}\right)^m = \frac{\left[-{\bar q}\right]^{m-l}}{\left[{\bar p}\left(1 - {\bar q} \right)\right]^m} \sum_{l=0}^m \, ^mC_l \, (b^\dagger_{(L)k})^l  \, (b_{(R)(-k)})^{m-l} \nonumber \\
\end{eqnarray}
Here, we have used the fact that the left and right Rindler operators commute with each other, that is,
\begin{eqnarray}
\left[(b^{(\, , \dagger)}_{(L)k},b^{(\, , \dagger)}_{(R)k}) \right] = 0
\end{eqnarray}
Combining the above equations together, the density matrix $\rho_{th}$ can finally be written completely in terms of the left and right Rindler operators and basis states as
\begin{eqnarray}
\rho_{th} &=&  \prod_k C_k^2 \sum_{m,p,q,l,l^\prime} \left[\frac{e^{-m  \beta^\prime \omega_k}  e^{-\frac{(p+q}{2} \beta \omega_k}\left[-{\bar q}\right]^{2m-l-l^\prime}}{\left[{\bar p}\left(1 - {\bar q} \right)\right]^{2m}}\right]\nonumber \\\nonumber \\
&& \times \left[\frac{\, ^mC_l \, ^mC_{l^\prime} \, ^pP_{m-l} \, ^qP_{m-l^\prime}}{m!p!q!} \right] \Theta_{p-m+l}\Theta_{q-m+l^\prime} \nonumber \\\nonumber \\
&& \times \sqrt{(p+l)!(q+l^\prime)!(p-m+l)!(q-m+l^\prime)!}\nonumber \\\nonumber \\
&& \times \; |_L (p+l)\rangle  \otimes |_R (p-m+l)\rangle \langle (q-m+l^\prime) |_R \otimes \langle (q+l^\prime)|_L \nonumber \\
\label{thermalbath3}
\end{eqnarray}
Taking  the trace of $\rho_{th}$ over the right Rindler modes and performing straightforward algebraically manipulations, we get the required result quoted in the text. (A detailed calculation and other  physical aspects of $\rho_r$ are discussed in a separate paper \cite{San}.)


\begin{thebibliography}{100}

\bibitem{Davies}
P. C. W. Davies,  J. Phys. A \textbf{8}, 609 - 616 (1975).

\bibitem{Unruh}
W. G. Unruh,  Phys. Rev. D \textbf{14}, 870 (1976).

\bibitem{unruhdet}
W. G. Unruh, Phys. Rev. D \textbf{14}, 870 (1976); Bryce DeWitt, The Global Approach to Quantum Field Theory (Clarendon Press, Oxford, 2003); N. D. Birrell and P. C. W. Davies, Quantum Fields in Curved
Space (Cambridge University Press, Cambridge, England, 1982).

\bibitem{unruhrev}
L. C. B. Crispino, A. Higuchi, and G. E. A. Matsas, Rev. Mod. Phys. \textbf{80}, 787 (2008).

\bibitem{accthermalbath}
T. Padmanabhan and T.P. Singh, Phys. Rev. D \textbf{38}, 2457-2463 (1988).

\bibitem{accthermalbath2}
S. S. Costa and G. E. A. Matsas, Phys. Rev. D \textbf{52},  3466-3471 (1995). 

\bibitem{brus}
D. E. Bruschi, N. Friis, I. Fuentes and S. Weinfurtner. New J. Phys. \textbf{15}, 113016 (2013)


\bibitem{Smolin}
L. Smolin Class. Quantum Grav. \textbf{3}, 347 (1986).

\bibitem{Steeg}
G. V. Steeg and N. C. Menicucci, Phys. Rev. D \textbf{79}, 044027 (2009); G. Salton, R. B. Mann, N. C. Menicucci, arXiv:1408.1395.

\bibitem{agullo}
I. Agullo, J. Navarro-Salas, G. J. Olmo and L. Parker, New J. Phys. \textbf{12}, 095017 (2010)

\bibitem{deser} 
S. Deser and O. Levin, Class. Quant. Grav. \textbf{14}, L163-L168 (1997).

\bibitem{San} 
S. Kolekar, Phys. Rev. D \textbf{89}, 044036 (2014).
  
\end{thebibliography}
\end{document}